# A NOVEL ARCHITECTURE FOR NETWORK CODED ELECTRONIC HEALTH RECORD STORAGE SYSTEM

B.VENKATALAKSHMI*, S. SHANMUGAVEL
Velammal Engineering College, Venkatalakshmi.tifac@velammal.edu.in
Anna University, ssvel@annauniv.edu

*Abstract*

The use of network coding for large scale content distribution improves download time. This is demonstrated in this work by the use of network coded Electronic Health Record Storage System (EHR-SS). An architecture of 4-layer to build the EHR-SS is designed. The application integrates the data captured for the patient from three modules namely administrative data, medical records of consultation and reports of medical tests. The lower layer is the data capturing layer using RFID reader. The data is captured in the lower level from different nodes. The data is combined with some linear coefficients using linear network coding. At the lower level the data from different tags are combined and stored and at the level 2 coding combines the data from multiple readers and a corresponding encoding vector is generated. This network coding is done at the server node through small mat lab net-cod interface software. While accessing the stored data, the user data has the data type represented in the form of decoding vector. For storing and retrieval the primary key is the patient id. The results obtained were observed with a reduction of download time of about 12% for our case study set up.

*Key words*

Network coding, Electronic Health Record system, Distributed content storage, RFID, Linear network code

1. **Introduction**

The fundamental idea of Network Coding spreads its potential in various network performance metrics, in the last decade. The use of network coding for large scale content distribution to improve download time (Christos et al 2005) is the motivation for this work of network coded Electronic Health Record (EHR) Storage System. With this initiatives of network coded content distributions, researchers contributed for, network coding based distributed storage in Sensor Networks (Dimakis et al 2005), reliability improvement under server failures(Bo Chen et al 2010), and more like this.

Today's widespread use of social networks, logs, videos, mails etc., makes an exponential increase in data storage. Application specific data centers like Supply chain process of an Industry, Health Care systems of Hospitals, Academic records of Educational Institutions etc., are also require maintenance of large volume of data. To obtain more reliable data storage for such cases, distributed data storage becomes essential. Distributed file storage in internet uses large data centers like Ocean store (S. Rhea et al 2001), Total Recall (R. Bhagwan et al 2004) etc. The entire system design requires offering reliable access to data.

Nowadays machine generated data exceeds human generated data and we need storage systems of the order of Exabyte's. When we consider distribution of data in wireless networks, there exists challenges like wireless data rates, link failures and packet loss probability. Thus we need to develop strategies by deployment and integration of newer technologies. The desirable performance metrics includes rebuild time, read/write bandwidth and storage efficiency. The major pulling factor is the tradeoff between reliability and redundancy. Intelligent architectures are required to achieve this and one such effort is this work by focusing on the metric rebuild time.

We make use of content based network coding for a selected scenario of EHR, which results in better storage and retrieval of contents. The rest of the paper is organized as follows. Section 2 reveals the related works and the basic principle used in our work. In section 3 we propose a new architecture for the design of network coded EHR-SS.

2. **Related Works**

Network Coding allows more intelligence at the nodes to perform simple computation (encoding). The data packets are combined and stored for distributed storage. Also the profit of network coding is achieved using linear transformations. Various research works have illustrated the benefits of Network Coding and the design of network coding for wired and wireless networks. The challenge in distributed storage arises when both the data sources and source nodes are distributed. A survey on the usage of network coding for distributed storage in wireless sensor networks (Alexandros G. Dimakis, Kannan Ramchandran 2008) paved a new way of research in the application of network coding. Network Coded Distribution storage systems with storage and repair bandwidth tradeoff is another milestone in the use of network coding.

File download time of almost 20-30% improvement is achieved (Christor Gkantsidis and Publo Rodriguez Rodriguez 2005) by the use of network coding for large scale content distribution. The results are tested in heterogeneous networks. To illustrate how network code improves the propagation of information without a global coordinated scheduler we consider the following (simple) example. In Figure 6.1 assume that Node A has received from the source packets 1 and 2. If network coding is not used, then, Node B can download either packet 1 or packet 2 from A with the same probability. At the same time that Node B downloads a packet from A, Node C independently downloads packet 1. If Node B decides to receive packet 1 from A, then both Nodes B and C will have the same packet 1 and, the link between them cannot be used.

If network coding is used, Node B will download a linear combination of packets 1 and 2 from A, which in turn can be used with Node C. Obviously, Node B could have downloaded packet 2 from A and then use efficiently the link with C.

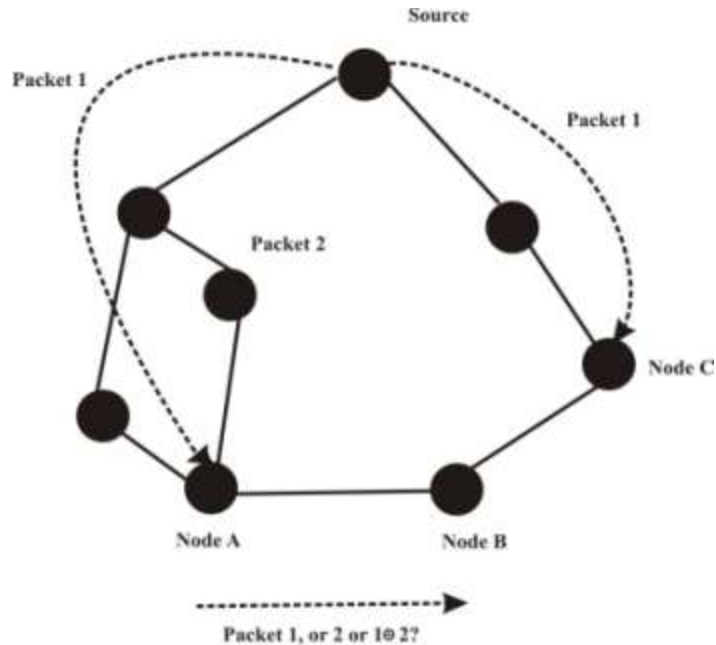

**Figure 2.1  Network Coding benefits**

## 3. Proposed Architecture For EHR-SS

Effective utilization of EHRs demands convergence of technologies, adoption of smart telephones and social media (Aviv Shachak and Alejandro R. Jadad 2010). Sue Ansell (2011) predicted the need for provincial licensing bodies for the tools of Social Networks. Followed by this, Health Care Networks evolved only as a source to share user's queries as a general forum. This capacity can be expanded for discovering the misuse of medical records as per the statements of You Chen (2011). Stanley Feld (2011) insisted on the need to practice Evident Based Medicine (EBM) which is a feasible practice by the merge of Social Media with Health Information Systems. The unexplored potentials of Social media concepts in diabetes self-management on mobile devices are highlighted by Taridzo Chomutare et al (2011).The above survey provide a thrust for our model of EHR-SS.

We propose a architecture of 4-layer to build the HER-SS as shown in Figure 3.1.  The lower layer is the data capturing layer.  This uses RFID passive tags.  The captured details are uploading on the clients in the 2$^{nd}$ layer. The networked clients are connected in this layer and they upload the data to the server in the 3$^{rd}$ layer. The network coded details exists in both 2$^{nd}$ and 3$^{rd}$ layer. The data cloud is optional and is constructs the 4$^{th}$ layer of the architecture.

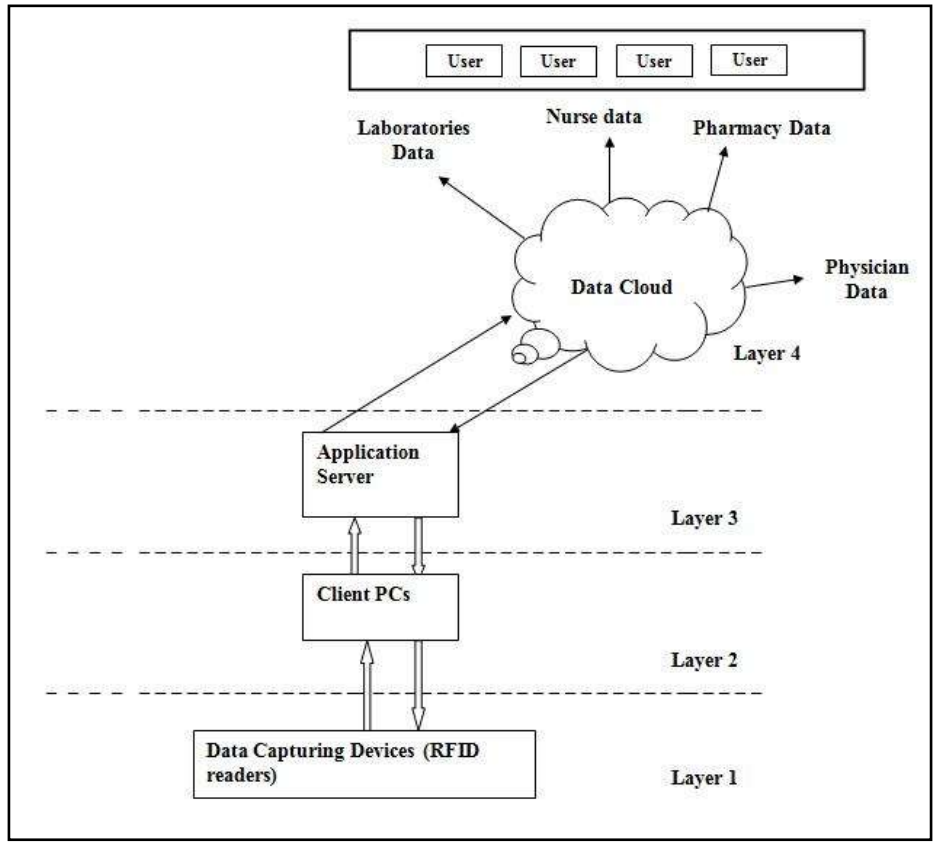

**Figure 3.1 Architecture of EHR-SS**

The network coding strategy used is as described in          and the notations used are described in Table 3.1.

**Table 3.1 Notations of Network coding model**

| S. No | Notation | Description |
|---|---|---|
| 1. | N | Number of tags |
| 2. | M | Number of Readers |
| 3. | K | Number of client nodes |
| 4. | $T_i$ | Tag ids; i = 1-N |
| 5. | $R_j$ | Reader ids; j = 1-M |
| 6. | $C_k$ | Client nodes; k = 1- K |
| 7. | {a,b,c} | Encoding coefficients –Level 1 Primary nodes |
| 8. | {p,q,r} | Encoding coefficients –Level 2 Primary nodes |
| 9. | {α, β, γ} | Encoding coefficients Secondary nodes |

## 3.1 RFID based Data Capture

The RFID system consists of a reader, tag and the host system. The reader and tag communicate through a RF signal link. Figure 3.2 shows components of a reader and the tag. The reader is the central nervous system of the entire RFID hardware system, establishing communication with the tag and the host system. The reader may be fixed (or table model) or handheld which is relatively costlier. The RFID tag is a device that can store and transmit data to a reader in a contactless manner using radio waves. RFID tags could be passive, active or semi-active (or semi passive). A passive tag has no on board power source and hence they have a limited range. They are also cheaper compared to an active or semi – active tag. The tag may also be a read only (RO), write once, read many (WORM) and Read – Write (RW) types. Both active and passive tags can be RO, or WORM or RW types. The RFID reader operates at the following frequencies, low frequency (125 KHz), high frequency (13.56 MHz), Ultra High Frequency (868 and 915MHz) and microwave frequency's (either 2.45 GHz or 5.8 Hz). A typical LF RFID system operates at 125 KHz or 134.2 KHz.

These RFID system have low data transfer rates from the tag to the reader and specially good if the operating environment contains metals liquids, dist, snow operating environment contains metals, liquids, dist, snow or mule. But the read range is low and need larger antennas refueling in higher cost tags further the tag memory capacity is also limited. However, they are least susceptible to performance degradation from metals and liquids.13.56 MHz is the typical frequency of use for HF RFID system. Compared to LF, HF tags are less expensive than LF tags. They also offer a fair performance in the presence of metals and liquids, HF RFIDS are currently the most widely available systems. The RFID system operating in these frequency ranges have the fastest data transfer rate between the tag and the reader. They are mostly meant for long range operation.

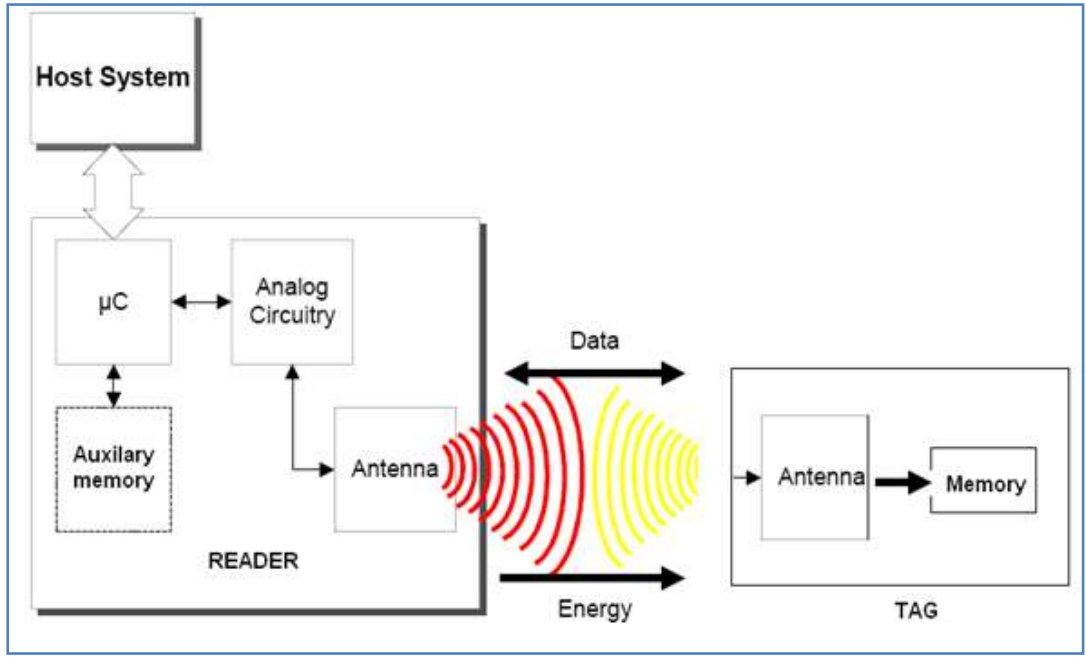

**Figure 3.2 Overview if RFID System**

## 3.2 Network Coding Strategy

The types of network coding have been discussed in section 1.8. This section applies a linear network coding procedure for distributed storage. The data captured in the lower level is from different nodes. The data is combined with some linear coefficients. There are 2-levels of linear coding as shown in Figure 3.3. The lower level combines the data from different tags and the level 2 coding combines the data from multiple readers. This network coding is at the client node. The higher layer network coding occurs at the secondary node or at the server.

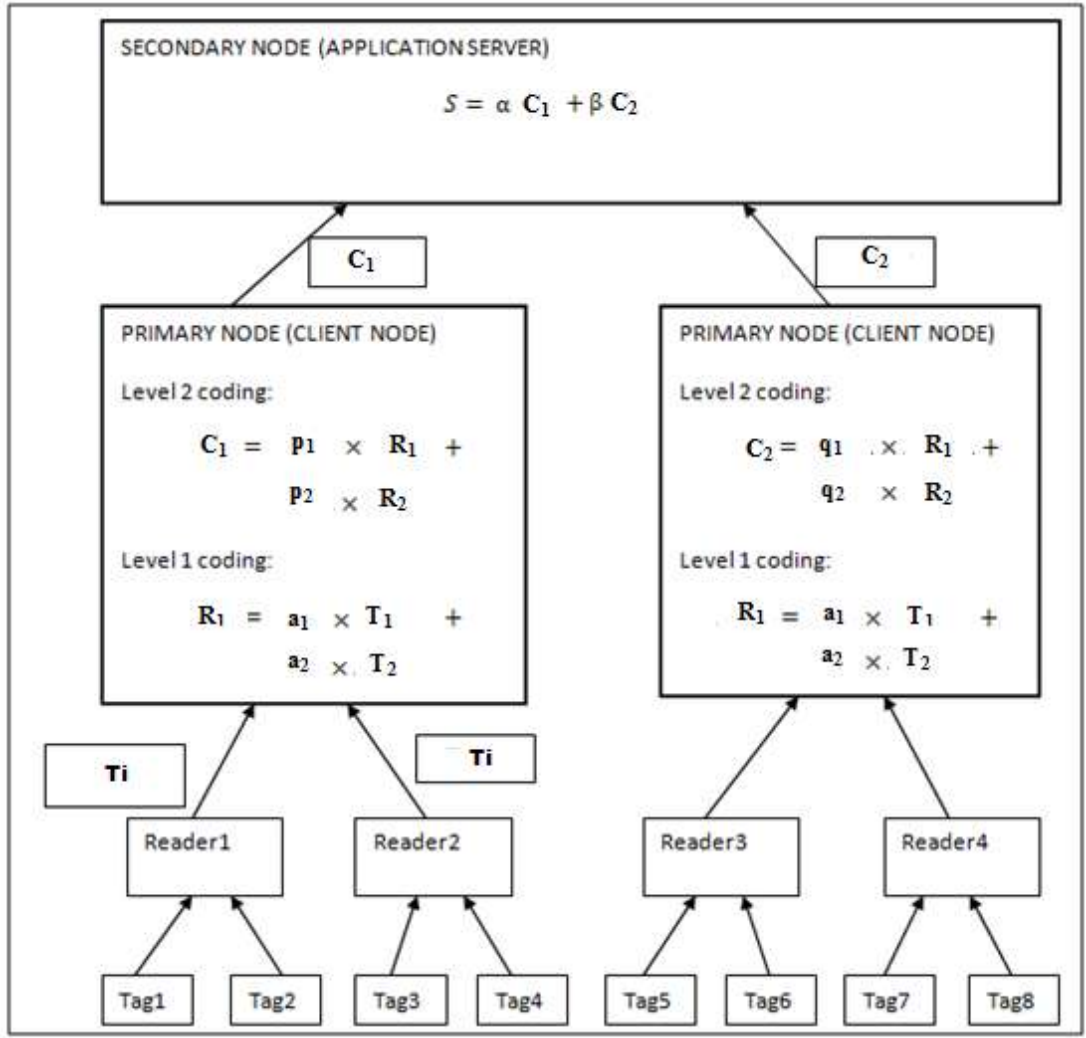

**Figure 3.3  Architecture for Multilevel network coding**

The strategical approach followed is described here. The data is captured through a .net application and the solution pushes the data to be combined into a network coding middleware. The middleware operation is performed in the matlab environment. The export and downcasting are the operations building the intraoperability.

The network coding middleware's abstracted view is shown in Figure 3.4.

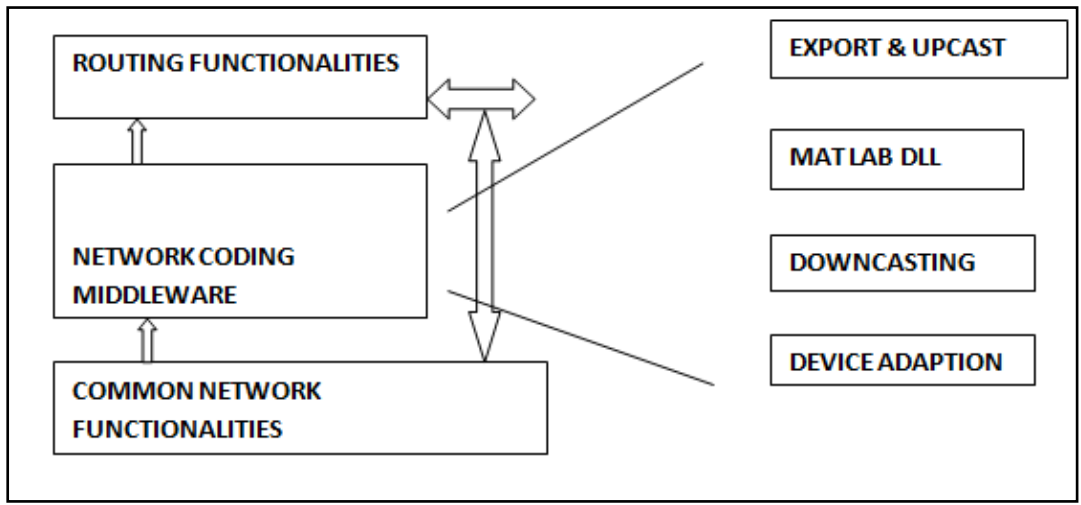

**Figure 3.4  Network coding middleware**

### 3.3   Client Server Model

The RFID readers are connected with the clients by Wi-Fi network model. The client support different mobile RFID readers. From the client the environment uses the local area network to connect the data to server. The hospital main server is visualized to get connected with many clients in the hospital environment.

## 4. CASE STUDY
### 4.1 EHR Scenario

The idea of our proposal addresses a Pervasive Hospital environment from where the data is uploaded to the cloud server. Also the data can be downloaded from the cloud to the user community. The overall architecture is modelled as a 4-layer set up and is depicted in Figure 3.1. The topmost or fourth layer forms the data cloud for the social media. The data cloud upload community in our experiment comprise of Physicians, Nurses, Lab technicians etc.  Whereas the download community includes the patients in the list and legal registration model is assumed for all the users. The data uploaded and downloaded is assumed to follow Health Level Seven (HL7) International format. The design of fourth layer of this overall architecture is beyond the focus of this work. Our work focuses on the experimental implementation of the first 3-layers. Layer 1 comprises of mobile RFID readers, layer 2 with client Personal Computers (PCs) and the 3$^{rd}$ layer with high end PCs acting as application servers. Our contributions in these layers are

1. RFID reader configuration and RFID healthcare application deployment in the first layer
2. Design and deployment of Network coding middleware in the primary and secondary nodes (client PCs and Application PCs respectively) of the second and third layers
3. Integration of the three layers and testing of upload and download processes

**4.2 Functional Blocks**

Handheld, Passive RFID readers are used as data capturing devices. The readers can be used to capture healthcare informations like Nursing details of In-ward patients, Observation reports of a physician, Lab report inferences, Hospital asset tracking etc., The modules tested in our work comprises of observation details of in-ward patients by the nurse and physician. The readers of mobile types are used. The devices follow passive EPC gen 2 protocols for tag communication. The captured data is exported to the second layer. The user module based applications are developed and deployed into the readers. Suitable admin module leverages the authenticated users to access their module. The authenticated users can import the details of the patients privileged to their limit. They can modify and export relevant observations about the patient.

The experiment uses Motorola's MC9090-G RFID handheld reader. The MC9090-G is a rugged mobile computer from Motorola which provides mobile workers with a flexible, always-on data connection to critical applications and systems. Equipped with the latest advances in mobile technology, the MC9090-G provides support for the richest enterprise applications, empowering mobile workers to capture and access critical and emergency information in real time. The MC9090-G offers the latest Intel processor designed to handle the specific demands of mobility, as well as robust persistent storage capabilities and multiple advanced data capture options. A choice of the two most robust Microsoft operating systems —Windows Mobile 5.0 or Windows CE 5.0 — gives the flexibility to select either a familiar feature rich environment or a robust customizable application specific environment.

The Figure 4.1 expresses the overview of the simple RFID system for inward patient. Each patient is provided with the wrist band where RFID tag is attached to it. The tag contains unique serial number which is the tag ID of the patient. Depending upon the memory capacity the tag may contain information like name of the patient, disease, etc. The nurse or the physician carries the RFID readers in order to find the patient who needs to medicate or to be observed periodically under various conditions.

The reader is configured either with USB or Bluetooth protocol. The reader facilitates Wi-Fi facility also but our experiment does not use the provision. The readers are synchronized with the help of Active Sync. We designed our application in a kiosk mode in each reader. The reader applications of various modules like registration (Admin module), medication (Nurse Module), observation/monitoring (Physician module) are developed using Visual studio. The application programming uses .net framework 3.5 with a database of SQLCE. A model of the application User interfaces are designed in PC and deployed in to the reader.

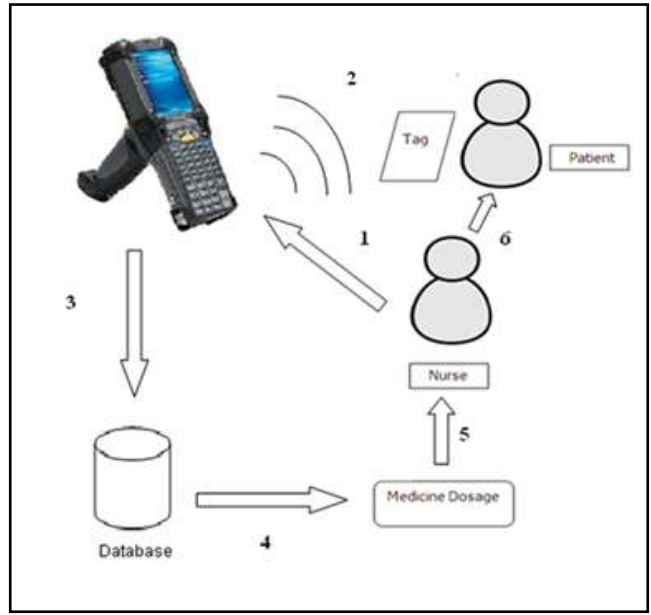

**Figure 4.1  RFID application set up**

The RFID readers capture the data from the tags and push them into the nodes of the upper layers. The contents comprise of different informations about the same patient like Medication and Lab inference or the informations about different patients. For the design of network coding middleware, we execute three sub-modules of design namely

    Module 1:  Modelling the network coding method
    Module 2: Network coding architecture at the nodes
    Module 3: Network Coding Middleware design

Module 1 of our design uses linear network coding. Each component of the information has its own encoding variable. The design follows the notations as listed in Table 3.1. Also Figure 3.3 depicts the model of linear coding followed in our work. Stage by stage content based encoding makes the retrieval of the contents comfortable. The primary nodes or the client nodes follow two level linear coding whereas the secondary nodes follow the single level linear coding. The complexity of the entire network coding processes depends upon the number of patients or tags, number of client systems and number of readers used.

Linear Coding at Client Nodes:

$$N = 2 \tag{4.1}$$

$$M = 2 \tag{4.2}$$

$$K = 2 \tag{4.3}$$

$$T_i = \{T_1, T_2\} \tag{4.4}$$

$$R_j = \{R_1, R_2\} \tag{4.5}$$

$$C_k = \{C_1, C_2\} \tag{4.6}$$

$$R_1 = a_1 \times T_1 + a_2 \times T_2 \tag{4.7}$$

$$R_2 = b_1 \times T_1 + b_2 \times T_2 \tag{4.8}$$

$$C_1 = p_1 \times R_1 + b_2 \times T_2 \tag{4.9}$$

$$C_2 = q_1 \times R_1 + q_2 \times R_2 \tag{4.10}$$

Server side encoding

$$S = \alpha \times C_1 + \beta \times C_2 \tag{4.11}$$

Module 2 consolidates the modified architecture of the nodes. The nodes are introduced with a new virtual layer of network coding. This middleware layer of network coding is sandwiched between the network common functionalities and the routing layers. The process of uploading accompanies encoding and the process of downloading involves level wise decoding. The functionality flow in the architecture is as follows. The Ethernet connected client stores the collected information from the reader. After that, the node seeks for forwarding the information. The demo set up has a virtual network layer for this purpose. The virtual network layer functions are grouped into 3-sublayers as common, network coding and routing. The common functionalities identify the common category of the patients or reader or clients. Then the middleware of network coding performs the linear network coding. The coded results are routed to the server with the routing sub layer functions.

Module 3 of our design provides the insight of our network coding middleware. The demo version of our middleware makes use of mat lab for linear coding. The entire middleware is developed as an application using .net. The application collects the tag informations in a string format and downcast the format. These down casted inputs are exported to mat lab. The common random encoding matrix generation and matrix multiplication forms the basic logic of our mat lab code. We pre-processed the information in the application and size it for (7,4) linear encode process. The mat lab code for linear encoding is compiled into a dynamic link library and patched with our .net application. The interoperability with mat lab is exploited in our demo version and the professional middleware can make use of server based format conversion and coding.

The informations collected by the reader are stored in the data base with suitable tables. After buffering them, the middleware of the virtual network layer search for the common tag id contents and the format conversion is executed. The information values are exported to Mat lab. The source code of linear netcod and the corresponding dlls contributed by Sadeghi, Shams and Traskovare (2010) are used as the basic blocks for our linear encoding. The scale down form of our code demo includes in the primary level an information matrix collected for a single patient. We assumed a 8*3 matrix, such that each column represents each module (Nurse module, Physician module and lab module), and the rows represent a 8-bit form of a data captured at the specific instance. This is a matrix assumed at a time stamp after one of the events and an observation period of 15-minutes. The encoding matrix is a binary matrix of size 3*1 at this level having each row value representing True or False about the event of each module occurrence. During decoding the request initiated by the corresponding module selection forms the encoding matrix. To avoid the code complexity in the demo simple data informations are coded and tested. The data storage and retrieval of network coded output is done with a local application server.

5. Output Samples

In the first stage of experiment, the application is deployed in the handheld reader. The .net based user interface as shown in Figure 5.1 allows the user to interact through the device. Synchronization of the device is essential and the output sample in Figure 5.2 reveals that. This invokes the application at the reader. The patient id is read as per sample in Figure 5.3. Figures 5.4,5.5,5.6, and 5.7 demonstrates the application sequence of medication.

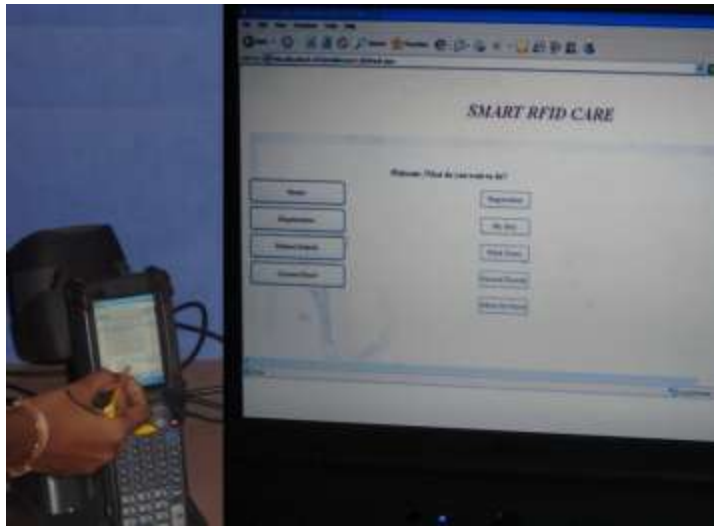

**Figure 5.1 User Interface**

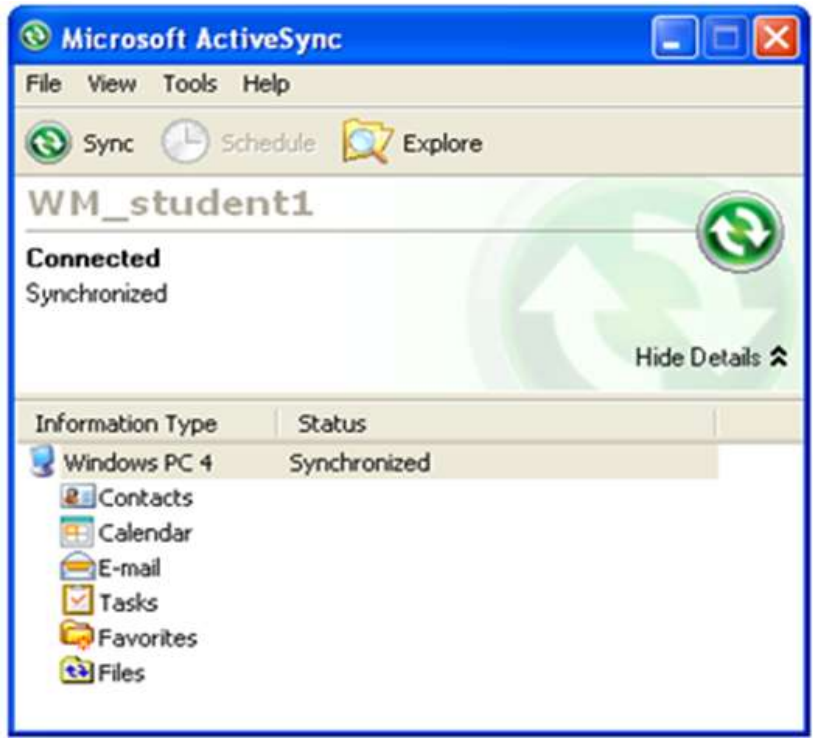

**Figure 5.2 Synchronization with RFID Reader**

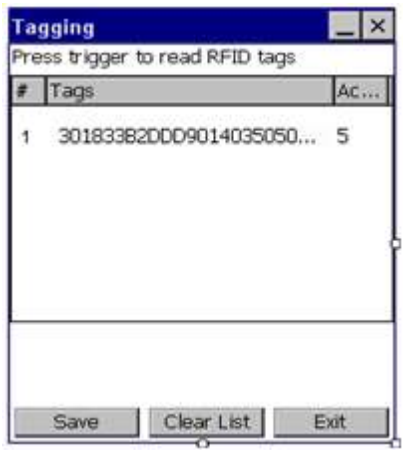

**Figure 5.3 Tag ID display**

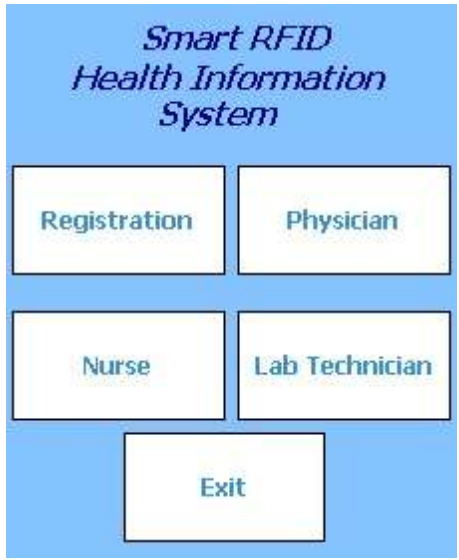

Figure 5.4  Registration Form

Figure 5.5  Patient details

**Figure 5.6  Case History**

**Figure 5.7 Medication details**

### 6. Conclusion

A simple linear network coding application is verified through our case study. The benefits of our design are (i) the server application is ignorant about the source of the contents collected at a specific time stamp (ii) while retrieving the data, the hand held device requests are very simple without specifying the details of their module; this saves the transmission cost (iii) In case of malfunction of any of the clients, the information can be retained with the history of encoding variables and the neighbour clients.